\documentclass[12pt]{article}
\usepackage{amsmath,epsf,amssymb,latexsym,enumerate,cite,shadow,array,color}
\usepackage{slashed}
\usepackage{graphicx,wasysym}
 \pdfoutput=1

\DeclareFontFamily{U}{rsf}{} \DeclareFontShape{U}{rsf}{m}{n}{
  <5> <6> rsfs5 <7> <8> <9> rsfs7 <10-> rsfs10}{}
\DeclareMathAlphabet\Scr{U}{rsf}{m}{n} \makeatletter
\@addtoreset{equation}{section} \makeatother

\def\be{\begin{equation}}
\def\ee{\end{equation}}
\def\ba{\begin{array}}
\def\ea{\end{array}}

\newcommand{\bea}{\begin{eqnarray}}
\newcommand{\eea}{\end{eqnarray}}

\def\rme{{\rm e}}

\usepackage{ifpdf}
\ifx\pdfoutput\undefined
   \pdffalse
\else
   \pdfoutput=1
   \pdftrue
  \usepackage[pdftex]{hyperref}
\newcommand{\rf}[1]{(\ref{#1})}

\begin{document}

\begin{titlepage}

\

\

\begin{center}
{\LARGE \textbf{
Non-minimal Inflationary Attractors\vskip 0.8cm }}

\

{\bf Renata Kallosh} and {\bf Andrei Linde}

{\sl Department of Physics and SITP, Stanford University, Stanford, California
94305 USA}\\
\end{center}
\vskip 1.5 cm

\

\begin{abstract}
Recently we identified a new class of (super)conformally invariant theories which allow inflation even if the scalar potential is very steep in terms of the original conformal variables. Observational predictions of a broad class of such theories are nearly model-independent. In this paper we consider generalized versions of these models where the inflaton has a non-minimal coupling to gravity with a negative parameter $\xi$ different from its conformal value $-1/6$. We show that these models exhibit attractor behavior. With even a slight increase of $|\xi|$ from $|\xi| = 0$, predictions of these models for $n_{s}$ and $r$ rapidly converge to their universal model-independent values corresponding to conformal coupling  $\xi = -1/6$. These values  of $n_{s}$ and $r$ practically coincide with the corresponding values in the limit $\xi \to -\infty$.
 \end{abstract}

\vspace{24pt}
\end{titlepage}



\newpage

\section{Introduction}

In  our recent papers \cite{Kallosh:2013pby,Kallosh:2013lkr,Kallosh:2013hoa,Ferrara:2013rsa} we further developed a superconformal approach to the descriptions of the broad class of models favored by observational data from WMAP9 \cite{Hinshaw:2012aka} and Planck2013 \cite{Ade:2013rta}. The superconformal approach to cosmology is based on our earlier papers with Kofman and  Van Proeyen \cite{Kallosh:2000ve} and more recent ones with  Ferrara,  Marrani and  Van Proeyen
\cite{Ferrara:2010yw}. The superconformal approach to supergravity in the form employed in \cite{Kallosh:2000ve} is described in detail in the recent book by Freedman and Van Proeyen \cite{Freedman:2012zz}, where one can find also the references to the original work.

The most surprising result of our recent investigation is the existence of  a new universality class of models of chaotic (super)conformal inflation \cite{Kallosh:2013hoa}. The new class of models found in  \cite{Kallosh:2013hoa} allows inflation even if the scalar potential is very steep in terms of the original conformal variables: Inflationary potential becomes exponentially flat upon switching to canonically normalized field variables in the Einstein frame.  In terms of the observational data, they all have an attractor point
\be
1 -n_{s} =2/N\, , \qquad r = 12/N^{2} 
\label{attractor}\ee
in the leading approximation in $1/N$, where $N$ is the number of e-folding of inflation. In \cite{Kallosh:2013lkr,Kallosh:2013hoa} we found that the model $R+R^{2}$ \cite{Starobinsky:1980te} also belongs to this broad  universality class, which explains why it has the same observational predictions as many other models of this class. Yet another class of superconformal cosmological attractors with closely related predictions was recently found in \cite{Ferrara:2013rsa}.

One may wonder whether these results may apply to the theories with non-minimal interaction of the scalar field to gravity described by the term ${\xi\over 2} \phi^{2}R$ with $\xi< 0$, for the case when $\xi$ differs from the conformal coupling $-1/2$. Ideally, one may try to implement this program in the context of (super)conformal theories, as we did before, but in this paper we will limit ourselves to a simpler task: We will investigate generalized models with $\xi< 0$ directly in the Jordan and Einstein frame. We will show that the models belonging to this new class also exhibit attractor behavior. With even a slight increase of $|\xi|$ from $|\xi| = 0$, predictions of these models for $n_{s}$ and $r$ rapidly converge to their universal model-independent values (\ref{attractor}) corresponding to conformal coupling  $\xi = -1/6$ . Moreover, we will show that in the leading approximation in $1/N$ these values  of $n_{s}$ and $r$ coincide with the corresponding values in the limit $\xi \to -\infty$.

One should note, that there is another class of models which exhibit a similar behavior, and leads to similar predictions: the models with potentials $~ \lambda\phi^{4}$ of a scalar field coupled to gravity with $\xi > 0$  \cite{Salopek:1988qh,Sha-1,Okada:2010jf,Bezrukov:2013fca,Kaiser:2013sna} and their supersymmetric generalizations   \cite{Einhorn:2009bh,Ferrara:2010yw,Lee:2010hj,Kallosh:2013pby}. The difference between these models with $\xi>0$ and the models with $\xi< 0$ studied in our paper is that the predictions of our new class of models with $\xi< 0$ are stable with respect to very strong modifications of the theory in the Jordan frame. 

In Section 2 we will describe the structure of the new class of models with spontaneously broken conformal symmetry, following \cite{Kallosh:2013hoa}, and then in Section 2 we will extend our results for the general coupling $\xi <0$.

\section{Chaotic inflation from conformal theory: T-Model}
In order to explain the basic idea of our approach, we will consider first a toy model with the following Lagrangian:
\begin{equation}
\mathcal{L} = \sqrt{-{g}}\left[{1\over 2}\partial_{\mu}\chi \partial^{\mu}\chi  +{ \chi^2\over 12}  R({g})- {1\over 2}\partial_{\mu} \phi\partial^{\mu} \phi   -{\phi^2\over 12}  R({g}) -{\lambda\over 36} (\phi^{2}-\chi^{2})^{2}\right]\,.
\label{toy}
\end{equation}
This
theory is locally conformal invariant under the following
transformations: 
\be \tilde g_{\mu\nu} = \rme^{-2\sigma(x)} g_{\mu\nu}\,
,\qquad \tilde \chi =  \rme^{\sigma(x)} \chi\, ,\qquad \tilde \phi =  \rme^{\sigma(x)}
\phi\ . \label{conf}\ee 
In addition, it has a global $SO(1,1)$ symmetry with respect to a boost between these two fields, preserving the value of $\chi^2-\phi^2$, which resembles Lorentz symmetry of special theory of relativity.

At the first glance, the physical interpretation of this theory may seem rather obscure. However, we may use the gauge $\chi^2-\phi^2=6$ and resolve this constraint in terms of the  canonically normalized field $\varphi$: 
$
\chi=\sqrt 6 \cosh  {\varphi\over \sqrt 6}$, $ \phi= \sqrt 6 \sinh {\varphi\over \sqrt 6} $.
Our action \rf{toy} becomes
\begin{equation}\label{chaotmodel1}
L = \sqrt{-g} \left[  \frac{1}{2}R - \frac{1}{2}\partial_\mu \varphi \partial^{\mu} \varphi -   \lambda \right].
\end{equation}
Thus our original theory is equivalent to a theory of gravity, a free massless canonically normalized field $\varphi$, and a cosmological constant $\lambda$  \cite{Kallosh:2013hoa}.

The main reason for doing this exercise was to show that the somewhat unusual term $(\phi^{2}-\chi^{2})^{2}$, or similar terms which will appear later in our paper, are essentially the placeholders to what will eventually look like a cosmological constant in the Einstein frame. The theories to be studied below are based on the idea that one can develop an interesting class of inflationary models by modifying this placeholder, i.e. by locally deforming the would-be cosmological constant. That is what we are going to do now.

Consider a  class of  models
\begin{equation}
\mathcal{L} = \sqrt{-{g}}\left[{1\over 2}\partial_{\mu}\chi \partial^{\mu}\chi  +{ \chi^2\over 12}  R({g})- {1\over 2}\partial_{\mu} \phi\partial^{\mu} \phi   -{\phi^2\over 12}  R({g}) -{1\over 36} F\left({\phi/\chi}\right)(\phi^{2}-\chi^{2})^{2}\right]\,.
\label{chaotic}
\end{equation}
where $F$ is an arbitrary function of the ratio ${\phi\over \chi}$. This theory is invariant under transformations (\ref{conf}), just as the toy model (\ref{toy}). When $F\left({\phi/\chi}\right)$ is constant, the theory has an additional $SO(1,1)$ symmetry, as we have seen in the example studied above. Introducing a function $F\left({\phi/\chi}\right)$ is the only possibility to keep local conformal symmetry (\ref{conf}) and to deform the $SO(1,1)$ symmetry.  The variable $z = {\phi/\chi}$ is the proper variable to describe the shape of the function $F\left({\phi/\chi}\right)$ in a conformally invariant way.

Using the gauge  $\chi^2-\phi^2=6$ immediately transforms the theory to the following equivalent form:
\begin{equation}\label{chaotmodel}
L = \sqrt{-g} \left[  \frac{1}{2}R - \frac{1}{2}\partial_\mu \varphi \partial^{\mu} \varphi -   F(\tanh{\varphi\over \sqrt 6}) \right].
\end{equation}
Note that asymptotically $\tanh\varphi\rightarrow \pm 1$ and therefore $F(\tanh{\varphi\over \sqrt 6})\rightarrow \rm const$, the system  in large $\varphi$ limit evolves asymptotically  towards its critical point  where the $SO(1,1)$ symmetry is restored.

It will be useful for us to  consider an alternative derivation of the same result, using the gauge $\chi(x) = \sqrt{6}$ instead of  the gauge $\chi^2-\phi^2=6$. The full
Lagrangian in the Jordan frame becomes
\begin{equation}
\mathcal{L}_{\rm total }= \sqrt{-{g_{J}}}\,\left[{  R({g_{J}})\over 2}\left(1-{ \phi^2\over 6}\right)-  {1\over 2}\partial_{\mu} \phi \partial^{\mu} \phi    -F\left(\phi/\sqrt 6\right) \left({\phi^2\over 6}-1 \right)^{2}\right]\,.
\label{toy2}
\end{equation}
Now one can represent the same theory in the Einstein frame, by changing the metric $g_J$ and $\phi$ to a conformally related metric $g_{E}^{\mu\nu} = (1- \phi^{2}/6)^{{-1}}  g_{J}^{\mu\nu}$ and a canonically normalized  field $\varphi$ related to the field $\phi$ as follows:
\be\label{field1}
 \frac{d\varphi}{d\phi} = {1\over 1- {\phi^2/ 6}} \ .
\ee
In new variables, the Lagrangian, up to a total derivative, is given by
\begin{equation}\label{LE1}
L = \frac{1}{2} \sqrt{-g} \left[ R - \frac{1}{2}g^{\mu\nu} \partial_\mu \varphi \partial_\nu \varphi -  V(\phi(\varphi)) \right],
\end{equation}
where the potential in the Einstein frame is 
\begin{equation}\label{eframe}
V(\phi) = {1\over 36} F\left(\phi/\sqrt{6}\right)  {\left(\phi^2-6 \right)^{2}\over \left(1-{ \phi^2\over 6}\right)^{2}} =  F(\tanh{\varphi\over \sqrt 6}) \ .
\end{equation}
This  brings the theory to the form   (\ref{chaotmodel}).

\begin{figure}[ht!]
\centering
\includegraphics[scale=1.1]{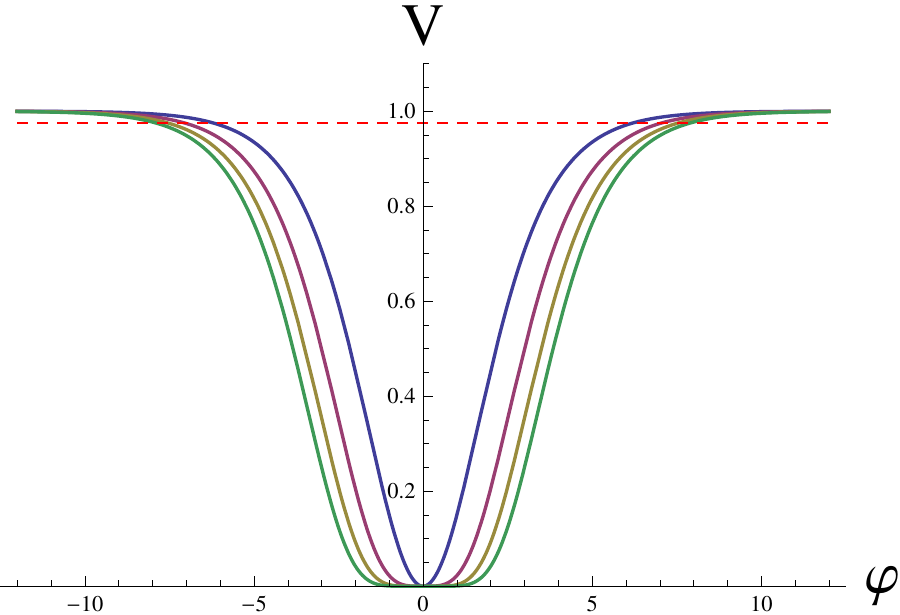}
\caption{Potentials of the T-Model  $V(\varphi) \sim {\tanh}^{2n}(\varphi/\sqrt6)$ for $n = 1,2,3,4$ (blue, red, brown and green,  corresponding to increasingly wider potentials).  These potentials differ from each other quite considerably, especially at $\varphi \lesssim 1$: at small $\phi$ they behave as $\varphi^{2n}$. Nevertheless all of these models predict the same values $n_{s} =1-2/N$, $r = 12/N^{2}$, in the leading approximation in $1/N$, where $N\sim 60$ is the number of e-foldings. The points where each of these potentials cross the red dashed line $V = 1-3/2N = 0.96$ correspond to the points where the perturbations are produced in these models on scale corresponding to $N = 60$. Asymptotic height of the potential is determined by the required amplitude of perturbations of metric; it is the same for all models of this class.}
\label{tmodelfig}
\end{figure}

We will start with a discussion of  the simplest set of functions $F\left({\phi/\chi}\right) = \lambda_{n} \left({\phi/\chi}\right)^{2n} = \lambda_{n} z^{2n}$. This is a generalization of the standard approach to chaotic inflation, where we originally used the simplest choice of functions $~\phi^{2n}$ \cite{Linde:1983gd}. Now we are making a similar choice, but in term of the homogeneous variables $z = \phi/\chi$, which preserve conformal invariance.
In this case one finds 
\begin{equation}\label{TModel}
V(\varphi) = \lambda_n\ {\tanh}^{2n}(\varphi/\sqrt6) .
\end{equation}
We called these models `T-Models'   because they originate from different powers of $\tanh(\varphi/\sqrt6) $,  and the potentials have the shape of the letter T, see Fig. \ref{tmodelfig}. Even though these potentials depend on $n$, observational predictions of these models do not depend on $n$.  Its basic representative $ \lambda_1\ {\tanh}^{2}(\varphi/\sqrt6)$ is the simplest version of the class of conformal chaotic inflation models proposed in \cite{Kallosh:2013hoa}. Rather unexpectedly, inflation is possible in a very broad class of models of this type, with very different potentials $F\left({\phi/\chi}\right)$ including very steep potentials $V(\phi)$ in terms of the Jordan frame field variable $\phi$. Moreover, almost all such models have nearly identical observational consequences, thus belonging to the same universality class \cite{Kallosh:2013hoa}. This is very different from what one could expect on the basis of investigation of inflation in random potentials, see e.g.  \cite{Marsh:2011aa}.

\begin{figure}[ht!]
\centering
\includegraphics[scale=0.36]{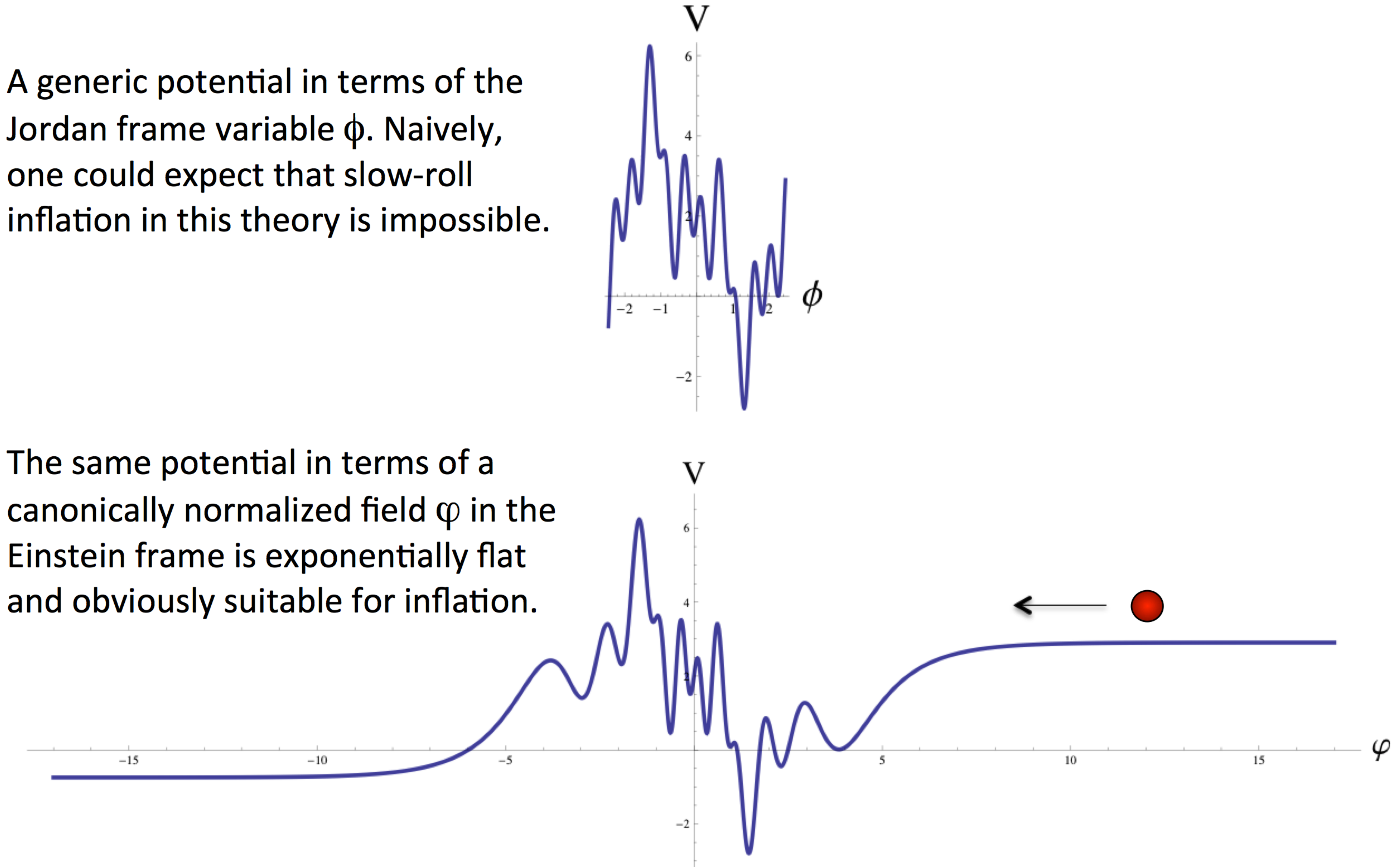}
\caption{Basic mechanism which leads to inflation in the theories with generic functions $F(\phi/\sqrt6)$. The potential can have a nearly arbitrary shape in terms of variables $\phi$. If this potential is non-singular at the boundary of the moduli space, it looks exponentially stretched and flat at large values of the canonically normalized field $\varphi$. This stretching  makes inflation very natural, and leads to universal observational predictions for a very broad class of such models \cite{Kallosh:2013hoa}. }
\label{stretch}
\end{figure}

To gain  intuitive understanding of these features of the new class of models, following  \cite{Kallosh:2013hoa}, let us consider an arbitrary non-singular potential $V(\phi) =F\left(\phi/\sqrt{6}\right)$ in terms of the Jordan frame variables, as shown in the upper panel of Fig. \ref{stretch}, and then plot the same potential in terms of the canonically normalized field $\varphi$ in the Einstein frame, shown in the lower panel of Fig. \ref{stretch}. As we see, the potential looks extremely flat upon switching to the canonically normalized field $\varphi$, which exponentially stretches and flattens the potential close to the boundary of the moduli space, sending this boundary to infinity, in terms of the field $\varphi$.\footnote{Other ways of flattening of the inflaton potentials were proposed in \cite{Dong:2010in}.}

\section{Models with arbitrary $\xi <0$}

Now we will consider a generalization of the Jordan frame representation of this class of models (\ref{toy2}), by replacing the conformal coupling $\xi = -1/6$ in  (\ref{toy2}) by a general coupling $\xi<0$:
The full
Lagrangian in the Jordan frame becomes
\begin{equation}
\mathcal{L}_{\rm total }= \sqrt{-{g_{J}}}\,\left[{  R({g_{J}})\over 2}\left(1+{\xi \phi^2}\right)-  {1\over 2}\partial_{\mu} \phi \partial^{\mu} \phi    - F\left(\sqrt{|\xi|}\phi\right) \left({\xi\phi^2}+1 \right)^{2}\right]\,.
\label{toy2a}
\end{equation}

One can present the same theory in the Einstein frame, by changing the metric $g_J$ and $\phi$ to a conformally related metric $g_{E}^{\mu\nu} = (1- \phi^{2}/6)^{{-1}}  g_{J}^{\mu\nu}$ and a new canonically normalized  field $\varphi$ related to the field $\phi$ as follows:
\be\label{field2}
 \frac{d\varphi}{d\phi} = {\sqrt{1+\xi\phi^{2}+6\xi^{2}\phi^{2}}\over 1+ {\xi\phi^2}} \ .
\ee
Just as before, in new variables, the Lagrangian in the Einstein frame is given by
\begin{equation}\label{LE12}
L = \frac{1}{2} \sqrt{-g} \left[ R - \frac{1}{2}g^{\mu\nu} \partial_\mu \varphi \partial_\nu \varphi -  F(\sqrt{|\xi|}\phi(\varphi))  \right].
\end{equation}
The main difference is that now  $\sqrt{|\xi|}\,\phi(\varphi)$ somewhat differs from $\tanh(\varphi/\sqrt 6)$. This difference is illustrated by the following set of figures.

\begin{figure}[ht!]
\centering
\includegraphics[scale=0.6]{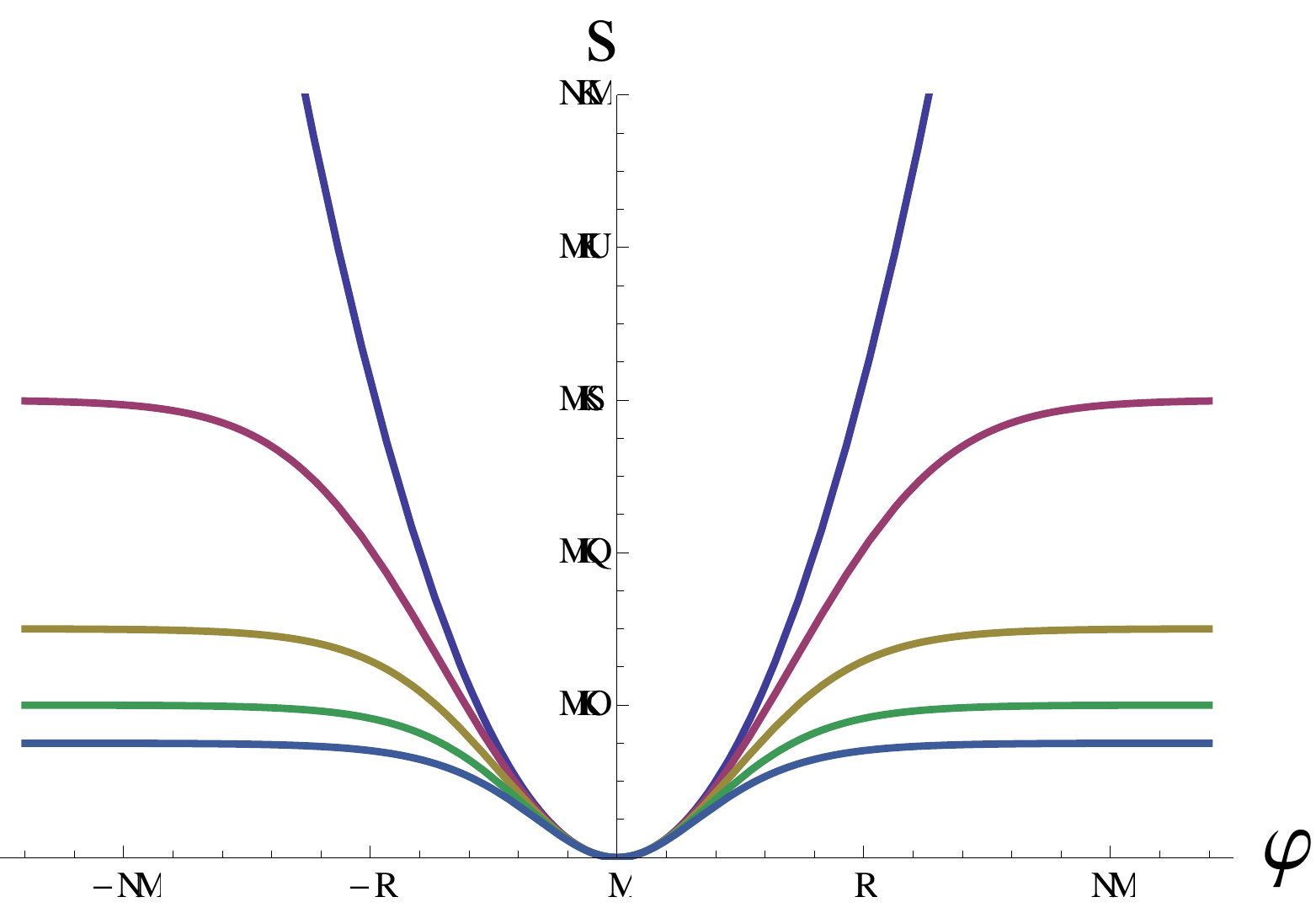}
\caption{The potential $V(\varphi)$ for $F(\sqrt{|\xi|}\phi) \sim \phi^{2}$. The upper curve corresponds to $\xi = 0$, other curves correspond to $\xi$ = -1/24, -1/12, -1/8 and -1/6. }
\label{tmodelfig1}
\end{figure}

The first one, Fig. \ref{tmodelfig1}, shows the potential $V(\varphi)$ corresponding to the simplest function $F(\sqrt{|\xi|}\phi) \sim \phi^{2}$. The series of images shows $V(\varphi)$ corresponding to a series of deferent parameters $\xi$, from $0$ to $-1/6$. For very small $|\xi|$, inflation occurs as in the chaotic inflation theory $\phi^{2}$ of the field $\phi$ minimally coupled to gravity. For larger $|\xi |$, inflation mostly occurs at the plateau of the potential $V$, and the results very rapidly approach the limiting form corresponding to $\xi \to -\infty$. This limit is almost reached when $\xi$ becomes smaller than $-10^{-1}$.

\begin{figure}[ht!]
\centering
\includegraphics[scale=0.6]{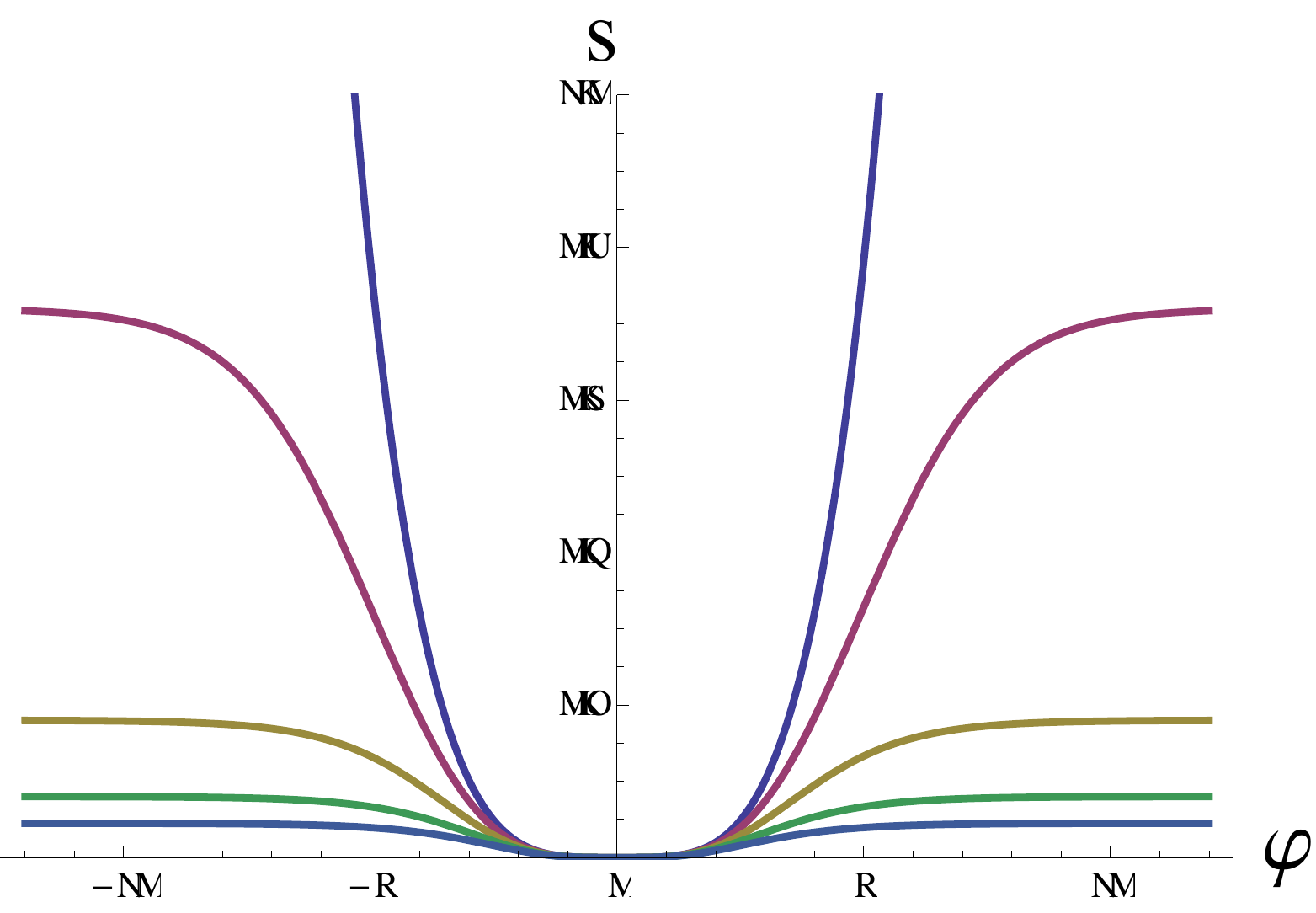}
\caption{The potential $V(\varphi)$ for $F(\sqrt{|\xi|}\phi) \sim \phi^{4}$. The upper curve  corresponds to $\xi = 0$, other curves correspond to $\xi$ = -1/24, -1/12, -1/8 and -1/6. }\label{tmodelfig2}
\end{figure}

Fig. \ref{tmodelfig2}, shows the potential $V(\varphi)$ corresponding to the $F(\sqrt{|\xi|}\phi) \sim \phi^{4}$. For very small $|\xi|$, inflation occurs as in the chaotic inflation theory $\phi^{4}$ of the field $\phi$ minimally coupled to gravity. Just as in the quadratic case, for larger $|\xi |$, inflation mostly occurs at the plateau of the potential $V$, and the results rapidly approach the limiting form corresponding to $\xi \to -\infty$. This limit is almost reached when $\xi$ becomes smaller than $-10^{-1}$.

\begin{figure}[ht!]
\centering
\hskip 0.5cm \includegraphics[scale=0.6]{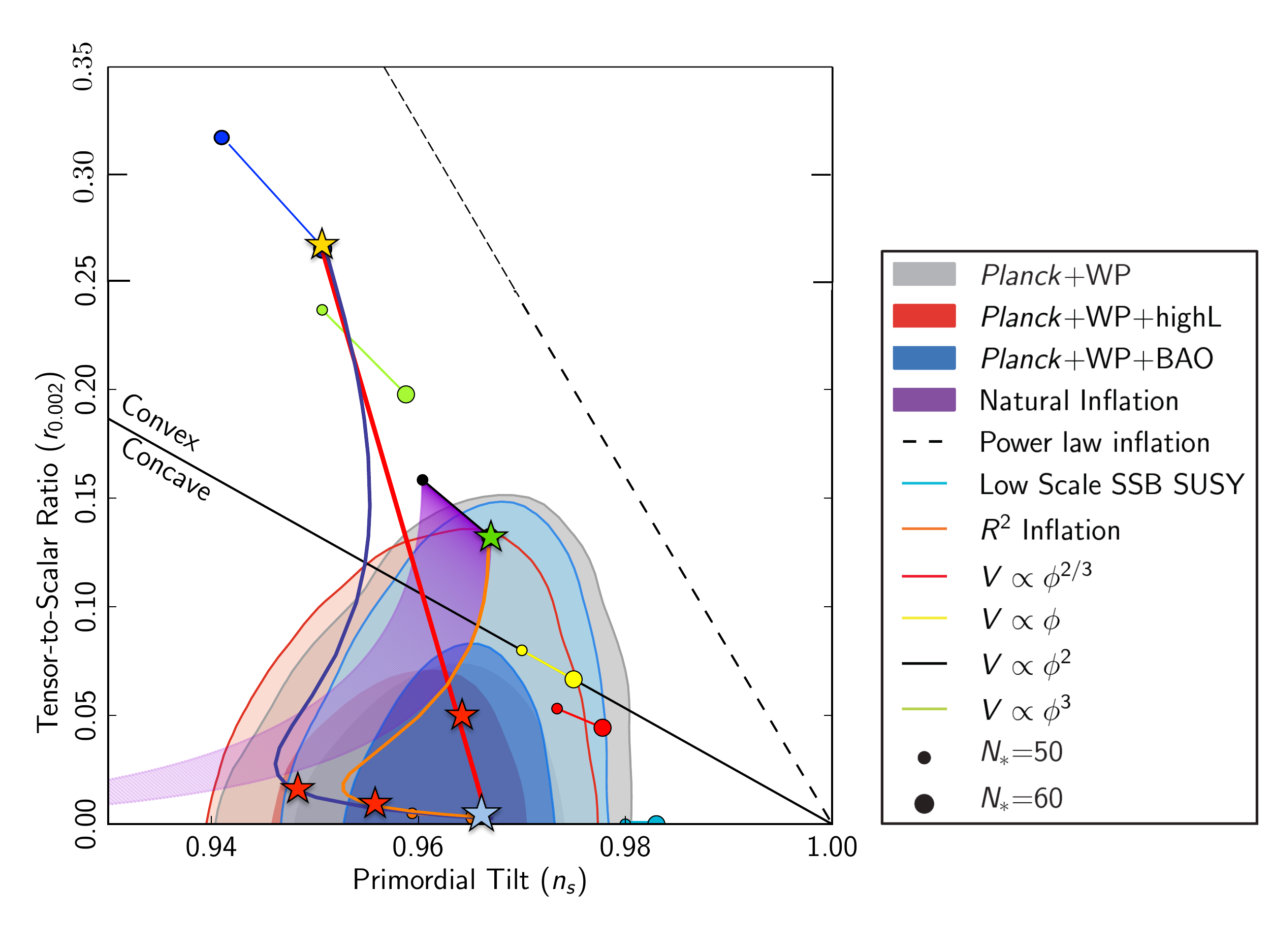}
\caption{Behavior of $n_{s}$ and $r$ in nonminimal T-Model with quadratic and quartic potentials as a function of the parameter $\xi$ in the interval from $0$ to $-1/6$, for the  number of e-foldings $N = 60$. The dark violet curve beginning with yellow star ($\xi = 0$) corresponds to the theory $\lambda\phi^{4}/4$. The orange curve beginning with the green star corresponds to the theory $m^{2}\phi^{2}/2$. The red line corresponds to the more conventional chaotic inflation in the theory $\lambda\phi^{4}/4$ with $\xi > 0$. In all of these theories, the point $\xi =0$ is a strong repeller: Even a tiny $\xi$ dramatically changes predictions of the theory. One can see it by looking at the red stars, which correspond to $\xi = -0.01$ for the T-model, and $\xi = 0.01$ for the usual chaotic inflation  with $\xi > 0$ (red line). Increase of $|\xi|$ beyond $1/6$ practically does not change predictions of the models. They reach at attractor shown by the blue star, which shows the unique prediction  of a broad class of models of  conformal chaotic inflation, including the T-model and the Starobinsky model \cite{Kallosh:2013hoa}.}
\label{tmodelfig3}
\end{figure}

The results of our investigation of inflationary predictions of these models are illustrated   in Fig. \ref{tmodelfig3}. As one can see, the blue star with $1 -n_{s} =2/N$, $r = 12/N^{2}$ is an attractor for the trajectories for all of these potentials.

What is the nature of this attractor behavior? Why did we found asymptotic saturation of observational predictions for $|\xi| \gtrsim 10^{-1}$. We already found the same attractor behavior for the broad class of models with conformal coupling $\xi = -1/6$, so let us try to understand what happens in the large $|\xi$ limit,  $|\xi| \gg 1/6$. 

Non-minimal coupling of the field $\phi$ can be ignored as long as $\xi\phi^{2} \ll 1$. However, the total number of e-foldings in the simplest model $m^{2}\phi^{2}/2$ minimally coupled to gravity  is
\be
N = {\phi^{2}\over 4} \ ,
\ee
so the value of the field $\phi$ corresponding to $N = 60$ is $\phi_{60} \approx 15$. In this case one can ignore non-minimal coupling only if $\xi\ll \phi_{60}^{-2} \sim 4\times 10^{{-3}}$. In the opposite limit, $\xi \gg  4\times 10^{{-3}}$, inflation mostly occurs near the boundary of the moduli space, where $1-\xi\phi^{2} \ll 1$. 
 In this limit, equation (\ref{field2})  for $\phi > 0$ acquires a particularly simple form:
\be\label{field3}
 \frac{d\varphi}{d\phi} = {\sqrt 6  |\xi|\phi\over 1-|\xi|\phi^{2}} \ .
\ee
The solution of this equation is
\be
\phi^{2} = {1\over |\xi|} \left(1-e^{-\sqrt{2\over 3}\varphi}\right) \ .
\ee
Thus  potentials originating from the function $F(\sqrt{|\xi|}\phi) \sim \phi^{2n}$ lead to a family of  inflationary potentials
\be
V(\varphi) \sim  \left(1-e^{-\sqrt{2\over 3}\varphi}\right)^{n} \ .
\ee
As explained in \cite{Kallosh:2013hoa}, all such potentials lead to the same observational predictions, which  coincide with predictions  in the first order in the inverse number of e-foldings $1/N$ with prediction of the large class of theories with conformal coupling $\xi  = -1/6$.

Under certain conditions, inflation may occur in the regime $1-\xi\phi^{2} \ll 1$ even for $\xi\ll \phi_{60}^{-2} \sim 4\times 10^{{-3}}$. Indeed, let us look again at Fig. \ref{stretch}. If the potential is very steep in the original Jordan frame variables, then one can have inflation only because of the stretching of the potential which occurs due to conversion to canonically normalized variable $\varphi$ near the boundary of the moduli space with $1-\xi\phi^{2} \ll 1$. Thus we are arriving to a rather interesting conclusion, which should be valid for generic potentials in the theories studied in  \cite{Kallosh:2013hoa} and in our paper: If inflation cannot happen in the central part of the moduli space because of the steepness of the  potential in terms of the field $\phi$,  it can happen only near the boundary of the moduli space, where inflationary predictions are robust and model-independent. In other words, we have a rather paradoxical situation: the more complicated and steep the potentials look in terms of the field $\phi$, the faster their inflationary predictions converge to the universal attractor point  $1 -n_{s} =2/N$, $r = 12/N^{2}$ shown in Figure \ref{tmodelfig3}. 

\section{Attractors and more attractors}
\begin{figure}[ht!]
\centering
\hskip 0.5cm \includegraphics[scale=0.5]{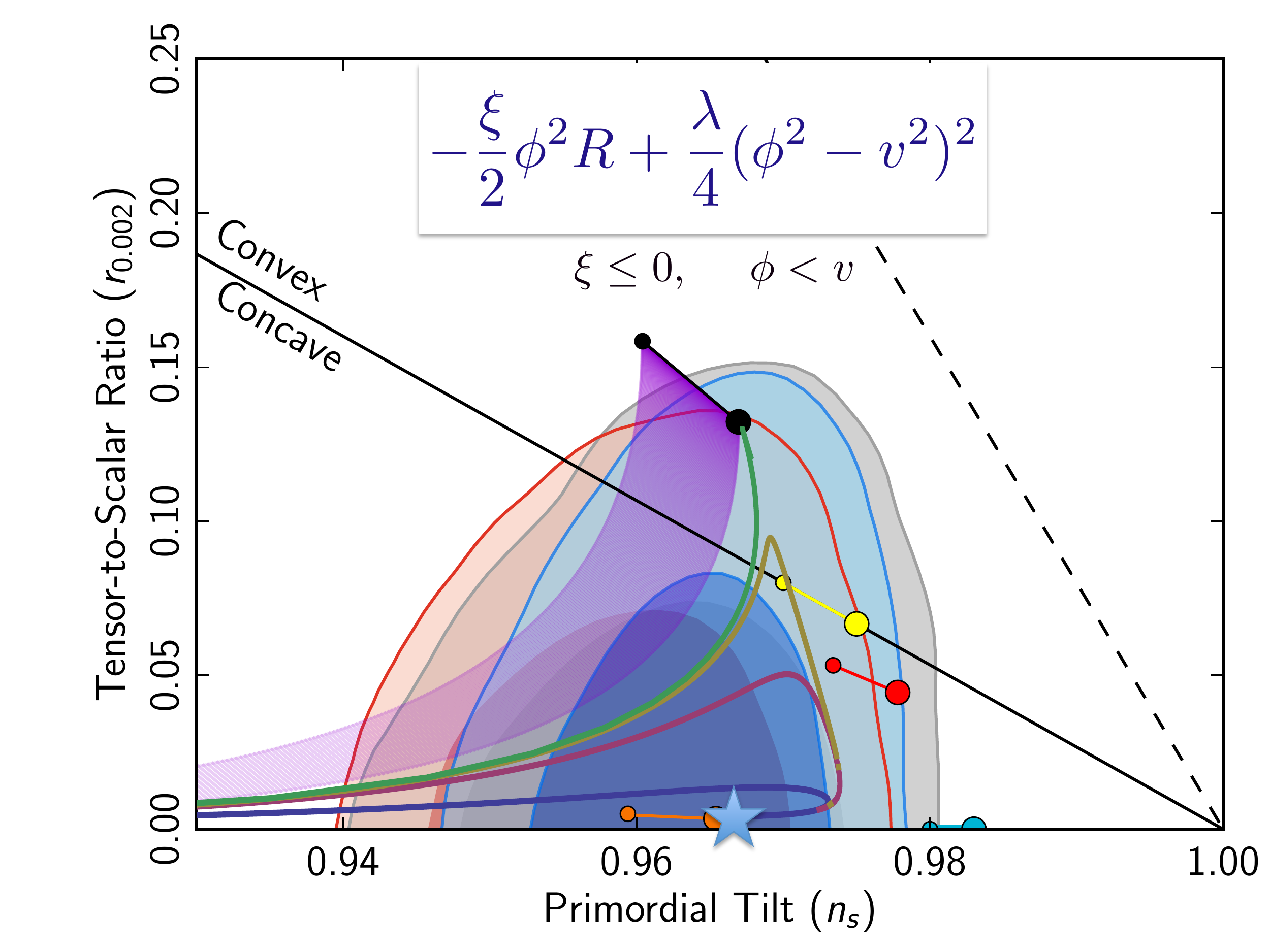}
\caption{Behavior of $n_{s}$ and $r$ in the Higgs model ${\lambda\over 4} (\phi^{2}-v^{2})^{2}$ for $\xi \leq 0$ with quadratic and quartic potentials as a function of the parameter $\xi$ in the interval from $0$ to $-1/10$, for the  number of e-foldings $N = 60$. The upper (green) line corresponds to the case $\xi = 0$. It grows up when $v$ increases. In the limit $v \gg 1$,  it reaches the point corresponding to the theory $m^{2}\phi^{2}/2$ \cite{Kallosh:2007wm}. As we see, this line comes closer to the center of the domain favored by Planck than the curves corresponding to natural inflation. The lower line corresponds to $\xi = -0.1$.  In the limit  when $v^{2} $ approaches $1/|\xi|$, all curves reach the same attractor point (\ref{attractor}) shown by a blue star.}
\label{tmodelfig4}
\end{figure}

This paper would be incomplete without mentioning another type of cosmological attractors which was found some time ago in a very similar context  \cite{Linde:2011nh}. Let us look again at the Jordan frame theory (\ref{toy2a}), and instead of the last term consider the simplest Higgs potential ${\lambda\over 4} (\phi^{2}-v^{2})^{2}$. Then the Einstein frame potential as a function  of $\phi$ for $\xi < 0$ will be 
\be
V(\phi)  = {\lambda\over 4}\ {(\phi^{2}-v^{2})^{2}\over (1-|\xi|\phi^{2})^{2}}= {\lambda\over 4\xi^{2}}\ {(\phi^{2}-v^{2})^{2}\over (\phi^{2}-{1\over |\xi|})^{2}}
\ee
One can consider two limiting cases here. For $v^{2} \ll {1\over |\xi|}$, this will be the usual Higgs potential, whereas  for $v^{2} = 1\over |\xi|$, this potential once again would represent a cosmological constant, just like in the theory (\ref{chaotic}) with $F = const$. Something interesting happens in the limiting case where $v^{2} = 1/|\xi| - \Delta$, where $\Delta \ll 1/|\xi|$.  If $\Delta$ is small enough, the potential remain almost constant for a broad range of $\phi$,  the narrow minimum of the Higgs potential appears in an immediate vicinity of the boundary of the moduli space. Therefore one can use equation (\ref{field3}) for investigation of stretching of the potential near the minimum of the potential. As a result, if one studies parameters $n_{s} $ and  $r$ as a function of $\Delta$, one finds that for sufficiently small  $\Delta$ the values of these parameters once again approach the attractor values (\ref{attractor}). This result, which is valid for any of $\xi < 0$ in the limit when $v^{2} $ and $1/|\xi|$ nearly coincide, was obtained in \cite{Linde:2011nh}. In Figure \ref{tmodelfig4}  we present the results of  \cite{Linde:2011nh} superposed with the Planck data.

\section{Conclusions}
In this paper we described a new class of inflationary models which exhibits the same cosmological attractor mechanism as the broad class of models based on spontaneously broken conformal and superconformal invariance discovered in \cite{Kallosh:2013hoa}. These developments are quite intriguing. As we see, these models share the same attractor point with the Higgs inflation  models \cite{Salopek:1988qh,Sha-1,Okada:2010jf,Bezrukov:2013fca} and their supersymmetric generalizations   \cite{Einhorn:2009bh,Ferrara:2010yw,Lee:2010hj,Kallosh:2013pby}, with the theory $R + R^{2}$ \cite{Starobinsky:1980te}, and with a class of theories with $\xi < 0$ studied in   \cite{Linde:2011nh}. From the purely theoretical perspective, one may recall that the discovery of the black hole attractor mechanism \cite{Ferrara:1995ih} was a starting point of a long series of fruitful investigations in supergravity and string theory. From the point of view of observational cosmology, these new developments make a strong case for the continuation of the hunt for the tensor modes all the way down to the level $3\times 10^{-3}$, as suggested by the predictions of this broad class of models.

This work was supported by the SITP and by the 
NSF Grant No. 0756174.  Part of this work was done during the School   ``Inflation and CMB Physics'' at Bad Honnef and the School ``Post-Planck Cosmology'' at Les Houches. We are grateful to the organizers of these schools for their hospitality.

\end{document}